

\input harvmac
\noblackbox
\pageno=0\nopagenumbers\tolerance=10000\hfuzz=5pt
\line{\hfill CERN-TH.7331/94}
\vskip 36pt
\centerline{\bf A DIRECT TEST OF PERTURBATIVE QCD AT SMALL $x$}
\vskip 36pt\centerline{Richard~D.~Ball\footnote{*}{On leave
from a Royal Society University Research Fellowship.}
 and Stefano~Forte\footnote{\dag}{On leave
from INFN, Sezione di Torino, Italy.}}
\vskip 12pt
\centerline{\it Theory Division, CERN,}
\centerline{\it CH-1211 Gen\`eve 23, Switzerland.}
\vskip 36pt
{\medskip\narrower
\ninepoint\baselineskip=9pt plus 2pt minus 1pt
\lineskiplimit=1pt \lineskip=2pt
\def\smallfrac#1#2{\hbox{${{#1}\over {#2}}$}}
\centerline{\bf Abstract}
\noindent
We show that recent data from HERA on the proton structure function
$F_2$ at small $x$ and large $Q^2$ provide a direct confirmation of the
double asymptotic scaling prediction of perturbative QCD. A
linear rise of $\ln F_2$ with the scaling variable $\sigma$ is observed
throughout the kinematic region probed at HERA, and  the measured
slope is in excellent agreement with the QCD prediction.
This provides a direct determination of
the leading coefficient of the beta function.
At large values of the  scaling variable $\rho$ the data
display a small but statistically significant scaling violation.
}
\vskip 16pt
\centerline{Submitted to: {\it Physics Letters B}}
\vskip 20pt
\line{CERN-TH.7331/94\hfill}
\line{June 1994\hfill}

\vfill\eject
\footline={\hss\tenrm\folio\hss}


\def\frac#1#2{{{#1}\over {#2}}}
\def\half{\hbox{${1\over 2}$}}

\def\smallfrac#1#2{\hbox{${{#1}\over {#2}}$}}

\def\GeV{{\rm GeV}}

\catcode`@=11 
\def\slash#1{\mathord{\mathpalette\c@ncel#1}}
 \def\c@ncel#1#2{\ooalign{$\hfil#1\mkern1mu/\hfil$\crcr$#1#2$}}
\def\lsim{\mathrel{\mathpalette\@versim<}}
\def\gsim{\mathrel{\mathpalette\@versim>}}
 \def\@versim#1#2{\lower0.2ex\vbox{\baselineskip\z@skip\lineskip\z@skip
       \lineskiplimit\z@\ialign{$\m@th#1\hfil##$\crcr#2\crcr\sim\crcr}}}
\catcode`@=12 

\def\PR{{\it Phys.~Rev.~}}

\def\NP{{\it Nucl.~Phys.~}}

\def\PL{{\it Phys.~Lett.~}}

\def\SPJETP{{\it Sov.~Phys.~J.E.T.P.~}}

\def\vol#1{{\bf #1}}\def\vyp#1#2#3{\vol{#1} (#2) #3}


\nref\DAS{R.D.~Ball and S.~Forte, preprint CERN-TH.7265/94, {\tt
hep-ph/9405320}.}
\nref\ZEUS{ZEUS~Collab.,\PL\vyp{B316}{1993}{412}.}
\nref\Hone{H1~Collab., \NP\vyp{B407}{1993}{515}.}
\nref\NMC{NMC~Collab., \PL\vyp{B295}{1992}{159}.}
\nref\ZEUSnew{G.~Wolf, talk at the International Workshop on Deep
Inelastic Scattering. Eilat, Israel, February 1994; M.~Roco, talk at the 29th
Rencontre de Moriond, March 1994.}
\nref\Honenew{K.~M\"uller, talk at the 29th Rencontre de Moriond, March 1994.}

Perturbative QCD predicts that at sufficiently large
$t\equiv\ln Q^2/\Lambda^2$ and small $x$ the nucleon structure
function $F_2$ should exhibit double scaling in the two variables
\eqn\esr{\sigma\equiv\sqrt{\ln\smallfrac{x_0}{x}\ln\smallfrac{t}{t_0}},
\qquad\rho\equiv\sqrt{\ln\smallfrac{x_0}{x}\big/\ln\smallfrac{t}{t_0}},}
provided only that the nonperturbative input to the perturbative
evolution is sufficiently soft. We have shown \DAS\ that this
prediction is indeed confirmed by the first measurements
of $F_2^p$ performed at HERA\refs{\ZEUS,\Hone}.
In fact, it turns out that not only most
of the HERA data, but even some of the older data from the NMC \NMC, lie
well inside the asymptotic regime, suggesting that the starting scale
$t_0\equiv\ln Q_0^2/\Lambda^2$ for the perturbative evolution should
be little more than $Q_0^2\sim 1~\GeV^2$. A significantly enlarged set
of measurements of $F_2^p$ has now  become available
\refs{\ZEUSnew,\Honenew}, which  makes it  possible to test double
scaling more quantitatively. Specifically, the slope of the linear
rise of $\ln F_2$ in the scaling variable $\sigma$ can  be
reliably measured, and turns out to be in excellent agreement with the
QCD prediction, thus giving a direct empirical determination of the
leading coefficient $\beta_0$ of the QCD beta--function.
We also find that there is now evidence for scaling violation at
large $\rho$.

\nref\DLLA{A.~De~Rujula, S.L.~Glashow, H.D.~Politzer, S.B.~Treiman,
           F.~Wilczek and A.~Zee,\hfil\break
\PR\vyp{D10}{1974}{1649}\semi
see also Yu.L.~Dokshitzer, \SPJETP\vyp{46}{1977}{641}.}
Double asymptotic scaling follows from a computation~\DLLA\ of the asymptotic
form of the structure function $F_2^p(x;t)$ at small $x$ based
on the use of the operator product expansion and renormalization group
at leading perturbative order. It thus
relies only on the assumption that any increase in $F_2^p(x;t)$ at small $x$ is
generated by perturbative QCD evolution, rather than being
due to some other (nonperturbative) mechanism manifested by an
increase in the starting distribution $F_2^p(x;t_0)$.
The resulting asymptotic behaviour takes the form
\eqn\eFasymp{F_2^p(\sigma,\rho)\sim
    N f(\smallfrac{\gamma}{\rho})
      \smallfrac{\gamma}{\rho}\smallfrac{1}{\sqrt{\gamma\sigma}}
    \exp\left[ 2\gamma\sigma -\delta\big(\smallfrac{\sigma}{\rho}\big)\right]
     \big[1+O(\smallfrac{1}{\sigma})\big],}
where $\gamma\equiv 2\sqrt{N_c/\beta_0}$,
$\beta_0=\smallfrac{11}{3}N_c-\smallfrac{2}{3}n_f$,
$\delta\equiv
(1+\smallfrac{2n_f}{11N_c^3})\big/(1-\smallfrac{2n_f}{11N_c})$,
and the unknown function $f$, which depends on the details of the
starting distribution, tends to one for sufficiently
small values of its argument. $N$ is an a priori undetermined
normalization factor.

In \DAS\ we derived \eFasymp\ by noting that at small-$x$ the one
loop QCD evolution equations reduce to wave equations, which propagate
the parton distribution functions from their boundary values at
$t=t_0$ and $x=x_0$ to larger values of $t$ and smaller values of $x$.
Since the propagation is unstable, away from the boundaries an
exponential increase with $\sigma$ of the form \eFasymp\ inevitably
arises, provided only that the small-$x$ behaviour of the starting
distributions at $t_0$ is sufficiently soft (which in
practice means that if $f_s(x;t)$ is a singlet parton distribution
function, $x^{1+\lambda}f_s(x;t_0)\to 0$ as $x\to 0$ for any
$\lambda\lsim 0.2$). The behaviour \eFasymp\ is thus a rather
clean prediction of perturbative QCD, in so far as it is
independent  of the details of the (soft) nonperturbative
parton distributions which are input at $t_0$, provided  that
at small $x$
these conform to expectations based on Regge theory. The
asymptotic behaviour  can be shown \DAS\ to set in rather rapidly
as $\sigma$ increases in a region  not too close to the boundaries,
i.e. when $\rho$ is neither too large nor too small.

In order to compare the data for $F_2^p$ with the prediction
\eFasymp\ we rescale the measured values of $F_2$ by a factor
\eqn\RFp{R_F'(\sigma,\rho)=R\exp\big(\delta(\sigma/\rho)
+\half\ln\sigma+\ln(\rho/\gamma)\big),}
to remove the part of the leading subasymptotic behaviour
which can be calculated in a model independent way.\foot{The constant
rescaling factor $R$ may of course be chosen arbitrarily; here we
choose $R=8.1$, so that the normalization of the figures is the same
as in \DAS.}
Then $\ln \left[R_F' F_2\right]$ is predicted  to rise linearly with
$\sigma$, independently of $\rho$ (when $\rho$ is large), with slope
\eqn\slope
{2\gamma=12\big/\sqrt{33-6n_f/N_c}=2.4}
if $n_f=4$ as in the HERA kinematic range.
The model-dependent  subasymptotic behaviour due to the
function $f$ can be eliminated by cutting all points with
subasymptotically small $\rho$; the scaling analysis of Ref.[1]
(see fig.~2 below) suggests that we place the cut at
$\rho^2 = 2$.

All the available experimental data\refs{\NMC,\ZEUSnew,\Honenew} for
$F_2^p$ which pass this cut are plotted
in \fig\fslope{Values of
$R_F'F_2^p$ plotted against $\sigma$: diamonds are ZEUS data \ZEUSnew,
squares H1 data \Honenew, and  crosses are NMC data.
The best fit straight line is also shown.}.
The predicted linear rise in $\sigma$ is spectacularly confirmed,
providing clear evidence that in the
region $\sigma^2 >1$, $\rho^2 > 2$ the asymptotic
behaviour \eFasymp\ has set in. Indeed, the scaling actually sets in
rather precociously: even the NMC data down to
$\sigma\sim 0.7$ seem to be rising linearly, with possibly
an indication of a systematic normalization mismatch of around
10\% between the NMC and the HERA determinations of $F_2$.

Fitting a straight line to all 80 HERA points in
the plot yields  a $\chi^2$ of 66, and a gradient $2\gamma_{exp}=
2.37\pm 0.16$,
in perfect agreement with the QCD prediction eq.\slope.
Turning this into a measurement of the leading coefficient of the
beta--function gives (with $N_c=3$) $\beta_0=8.6\pm 1.1$ (to
be compared with 25/3 for $n_f=4$).
This is a direct, model independent, and highly nontrivial test of the
perturbative dynamics of asymptotically free nonabelian gauge theory.

We next consider  scaling violations, both in the subasymptotic region of
small $\sigma$ and small $\rho$, and in the post-asymptotic region of
large $\rho$. This is best done by rescaling $F_2^p$ by a factor
\eqn\RFp{R_F(\sigma,\rho)=R\exp\big(-2\gamma\sigma + \delta(\sigma/\rho)
+\half\ln\sigma+\ln(\rho/\gamma)\big)}
to remove all the leading behaviour in
\eFasymp. The rescaled structure function should thus scale in both
$\sigma$ and $\rho$ when both are sufficiently large to lie in the
asymptotic region: $R_F F_2^p = N +O(1/\sigma) + O(1/\rho)$. This
double asymptotic scaling behaviour is tested in the two
scaling plots \fig\fscaling{$R_F F_2^p$ plotted against a) $\sigma$ and
b) $\rho$. Included in the plots are all the HERA data with $\rho
>1.2$, $\sigma> 0.7$, respectively. The curves show the prediction
obtained \DAS\ evolving a typical
soft  starting gluon distribution: a) dot-dash curve, $\rho=1.4$; solid
curve,
$\rho=2.2$; dotted curve, $\rho=3.2$. b) dot-dash curve, $\sigma=1.1$; solid
curve,
$\sigma=1.8$; dotted curve, $\sigma=2.1$.},
where we also display the predictions obtained \DAS\ by applying the
leading small-$x$ form of the evolution equations to a typical soft
starting gluon distribution.
Specifically, \fscaling a) shows that the scaling in $\sigma$ sets in
very rapidly, as all the points on the plot lie in the asymptotic
regime; \fscaling b) shows that the scaling in $\rho$ only sets in for
$\rho^2\gsim 2$. However even if $\rho$ is as low as
$\rho\sim\half$ the subasymptotic corrections
due to $f(\gamma/\rho)$ seem fairly well accounted for  by the
scaling violation displayed by the curves of \fscaling.

\topinsert\hfil
\vbox{\tabskip=0pt \offinterlineskip
      \def\tablerule{\noalign{\hrule}}
      \halign to 350pt{\strut#&\vrule#\tabskip=1em plus2em
                   &\hfil#\hfil&\vrule#
                   &#\hfil&\vrule#
                   &#\hfil&\vrule#
                   &\hfil#&\vrule#\tabskip=0pt\cr\tablerule
      &&\omit&&\omit\hidewidth $N_s$\hidewidth
             &&\omit\hidewidth $N_h$\hidewidth
             &&\omit\hidewidth $\chi^2$\hidewidth&\cr\tablerule
   &&   a)  && $0.341\pm 0.005$ && 0               &&  96 &\cr
   &&   b)  && 0                && $0.156\pm0.002$ && 878 &\cr
   && a)+b) && $0.319\pm 0.012$ && $0.012\pm0.002$ &&  91 &\cr\tablerule}}
\hfil\bigskip
\centerline{\vbox{\hsize= 380pt \raggedright\noindent\footnotefont
Table: The fitted normalizations $N_s$ and $N_h$ and
the associated $\chi^2$s (103 data points).
The different cases considered are a) soft pomeron b)
hard pomeron, and the linear combination a) + b).
}}
\bigskip
\endinsert

More interestingly,
at large $\rho$ there now appears to be a statistically significant
rise above the  scaling prediction.
To test the significance of this rise, we fitted to
the data a linear combination of the behaviour discussed above and
displayed by the curves of \fscaling, and a ``hard pomeron'' behaviour,
which violates scaling by rising with $\rho$
(see ref.\DAS\ for a more detailed discussion).
Including in the fit the 103 HERA points with both $\sigma^2$ and
$\rho^2$ greater than one half gives the results  displayed in the
table. The data seem to prefer a $4\pm 1\%$ admixture of the hard
pomeron solution.
One should be very cautious about taking this as
evidence for the hard pomeron per se, however, since higher loop
corrections should give a similar rise\ref\EKL{R.K.~Ellis,
Z.~Kunszt, and E.M.~Levin, Fermilab-PUB--93/350-T, ETH-TH/93--41, to
be published in \NP B.}.
It should be possible to settle this issue decisively when
a more detailed set of data and more accurate theoretical calculations become
available.

{\bf Acknowledgement:} We are very grateful to Dick Roberts for
communicating to us the data of refs \refs{\ZEUSnew,\Honenew}.


\listrefs
\listfigs
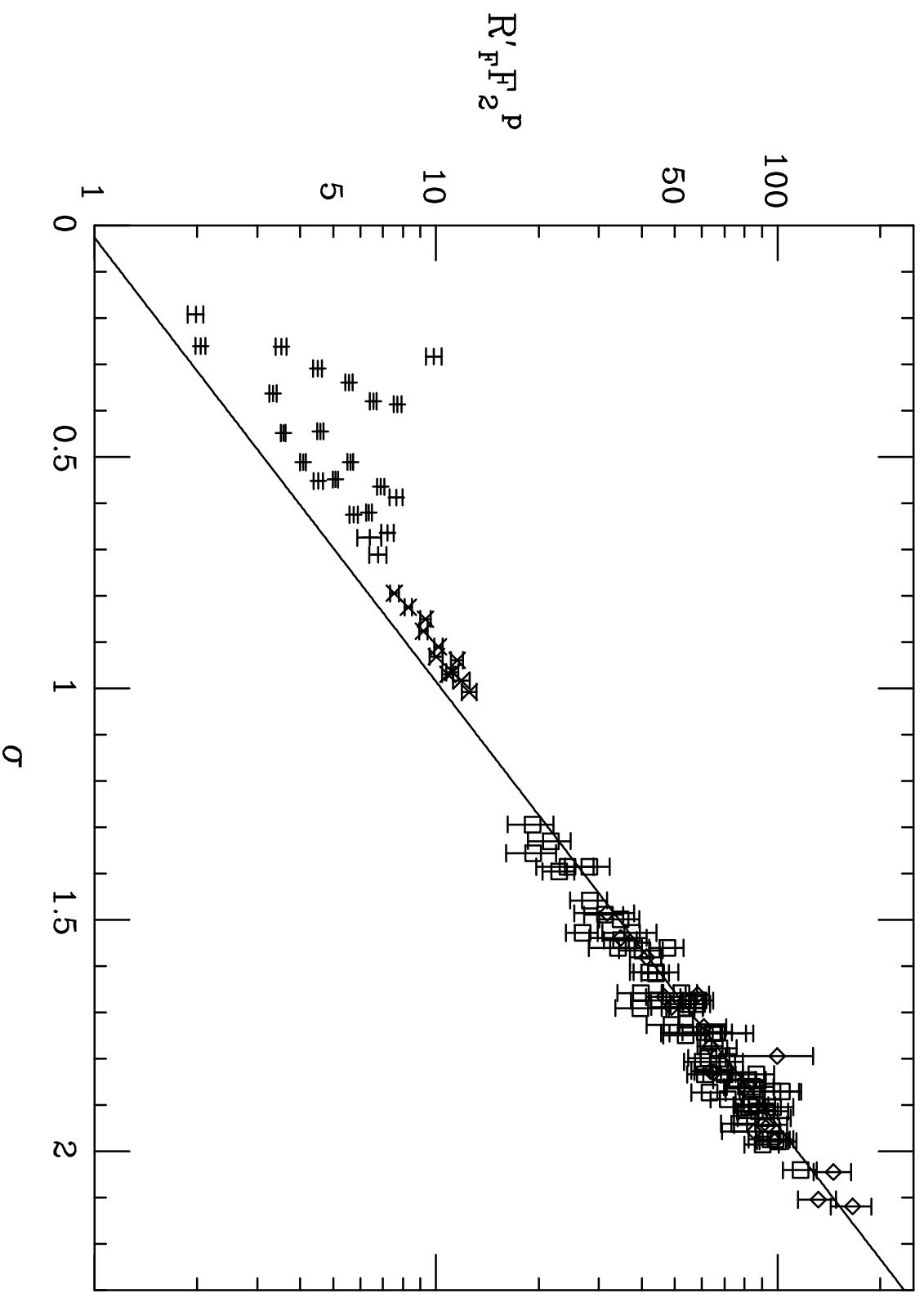
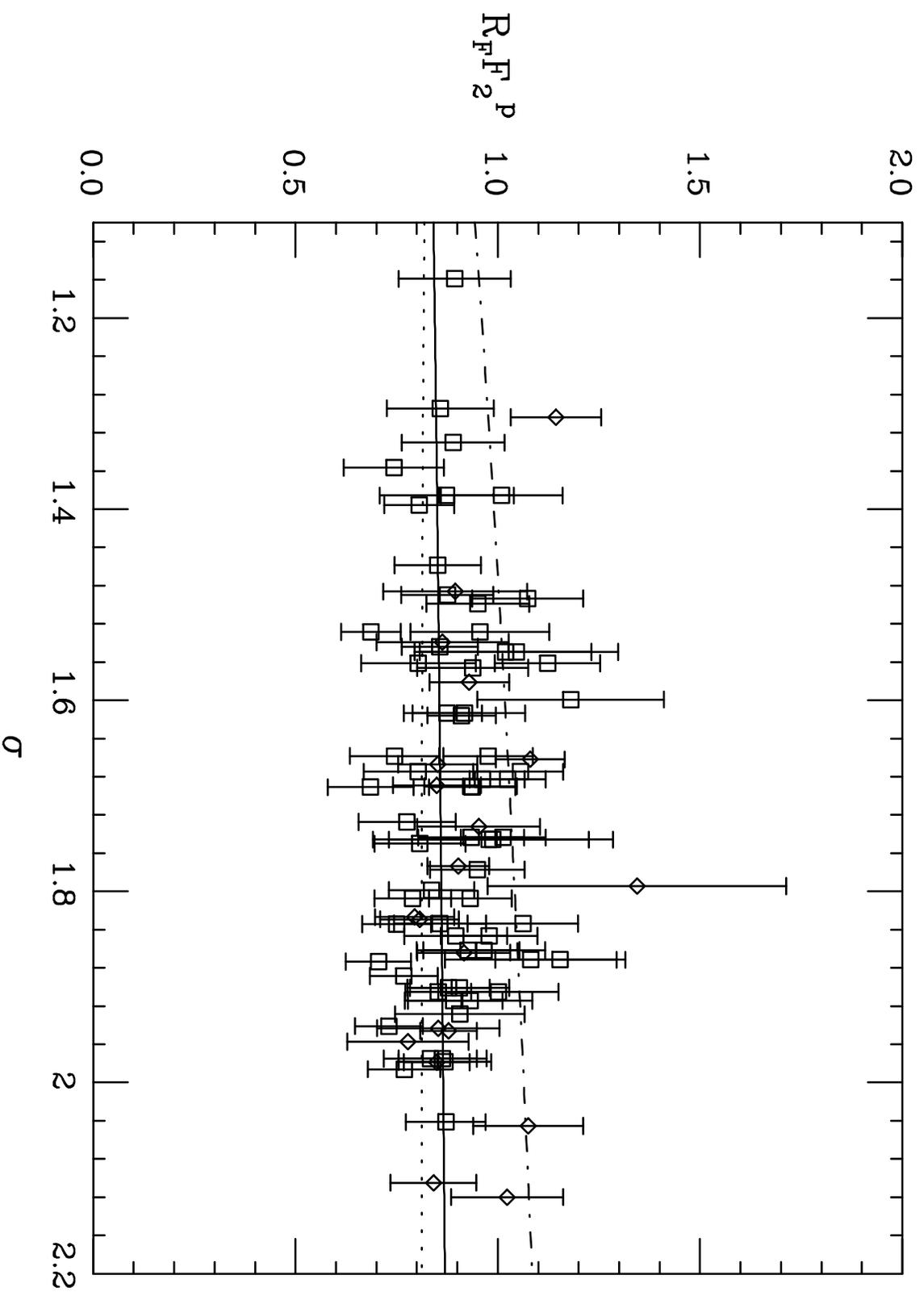
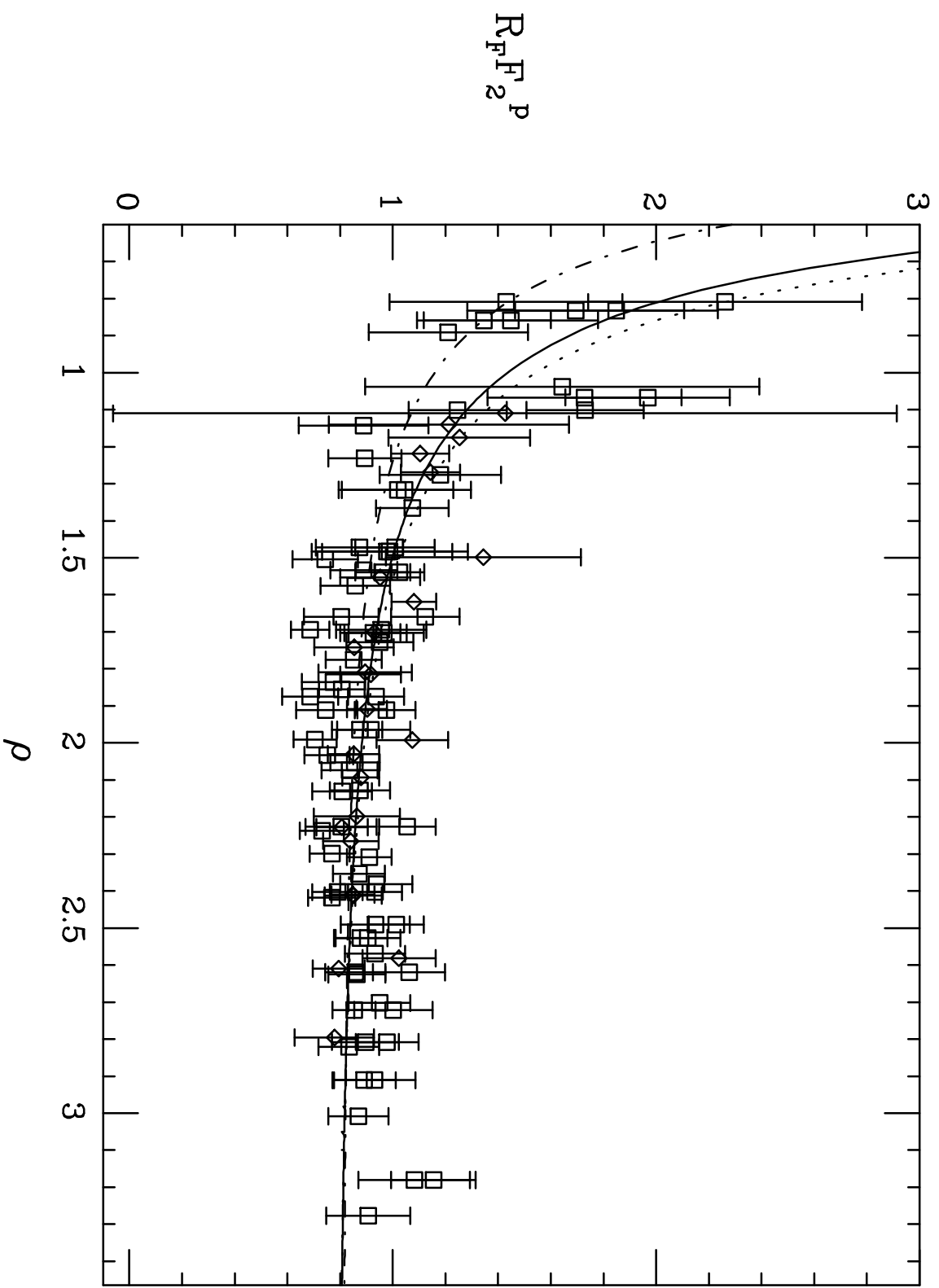
\end